# Spectral model of time-domain coherent anti-Stokes Raman scattering

**Michele Marrocco**

*ENEA, via Anguillarese 301, 00167 Rome, Italy*
michele.marrocco@enea.it

We show that the increasingly popular nonlinear optical technique of time-domain coherent anti-Stokes Raman scattering (CARS), which is usually understood in terms of the semiclassical time-dependent third-order polarization, can be equally explained in terms of the time-delayed version of the Yuratich equation so popular in traditional frequency-domain CARS. The method brings out the strong dependence of CARS time traces and time-delayed CARS lineshapes on the spectral envelope of the probe laser electric field. Examples are analytically shown for experimental results that are otherwise treated by means of numerical methods only.

Lately, time-resolved and time-delayed coherent anti-Stokes Raman scattering (CARS) has attracted a lot of interest as a diagnostic tool in applications to condensed media and gases [1-19]. Its success has been guaranteed by the commercial availability of laser systems that deliver optical pulses with picosecond (ps) and femtosecond (fs) durations so that the advantages of short and ultrashort excitation may be combined with the more typical advantages of chemical selectivity and spectral contrast that are common to frequency-domain spectroscopies. The use of such laser sources has, indeed, led to different experimental schemes (namely, ps-CARS [8, 10, 17], fs-CARS [1, 3, 5, 8, 19] and hybrid fs/ps CARS [2, 4, 6-9, 11-16, 18]) that overcome some of the drawbacks of traditional CARS realized with nanosecond (ns) laser pulses [8, 20, 21]. According to these schemes, the pump and the Stokes electric fields, whose synchronous envelopes are respectively $E_1(t)$ and $E_2(t)$, initiate the Raman response described by the time-dependent third-order susceptibility $\chi^{(3)}(t)$ that is later probed by means of a third field with envelope $E_3(t)$. The time delay between the pump/Stokes pulses and the probe pulse is such that two of the main disadvantages of traditional ns-CARS are minimized (when not made vanish!), namely the strong non-resonant contribution to the total signal and the intricate dependence on the collisional environment.

Although a lot of effort has gone into experimental advances, theoretical research insists on the dynamical approach within the semiclassical picture of Raman interaction [22] . In this context, the time-dependent third-order polarization that creates the CARS signal is calculated to be

$$P^{(3)}(t,\tau) = -E_3(t)\int_0^\infty d\tilde{t}\, e^{i(\omega_1-\omega_2)\tilde{t}} \chi^{(3)}(\tilde{t}) E_1(t+\tau-\tilde{t}) E_2^*(t+\tau-\tilde{t}) \quad (1)$$

where $\omega_1$ and $\omega_2$ are the frequencies of the pump and Stokes fields and $\tau$ is the time delay [22]. The CARS signal $S(\omega_{aS},\tau)$ at the anti-Stokes frequency $\omega_{aS}$ is finally obtained as the square modulus of the Fourier transform of Eq. (1) and, to ease the task, a fast electronic dephasing resulting in an exponential decay for the susceptibility is ordinarily assumed on reasonable grounds [1, 2, 6, 8, 10, 11, 13-18, 22].

Here, we follow a different route. Instead of pursuing the time model of Eq. (1), which is persistently invoked to illustrate experimental data [1-3, 6, 8, 11, 15, 18, 22], we use the spectral version of Eq. (1) containing the frequency envelopes $\mathcal{E}_j(\omega_j)$ (with $j$ = 1, 2, 3) of the fields. This spectral model was first suggested by Yuratich for the calculation of CARS lineshape of an isolated Raman transition of linewidth $\Gamma$ (FWHM) [23] and, although it has been usually restricted to synchronous ns-laser pulses [24-26], some authors (not aware of Yuratich's work) have reintroduced the idea in the context of fs-CARS for coherent control by means of spectral phase shaping [4, 27]. Unlike this approach, we tailor the model to time-resolved information only (i.e., without spectral phase shaping). Thus, after some algebra, it is possible to write a general equation for the CARS polarization as a function of the time delay $\tau_{12}$ between the pump and Stokes pulses and the time delay $\tau_{23}$ between the Stokes and probe pulses. The result is the time-delayed Yuratich equation

$$P^{(3)}_{CARS}(\omega_{aS},\tau_{12},\tau_{23}) \propto \int_{-\infty}^{\infty}\int_{-\infty}^{\infty} e^{-i\omega_1\tau_{12}} e^{-i\omega\tau_{23}} \frac{\mathcal{E}_1(\omega_1)\mathcal{E}_2^*(\omega_1-\omega)\mathcal{E}_3(\delta_{aS}-\omega)}{\Delta-\omega-i\Gamma/2} d\omega_1 d\omega \quad (2)$$

with $\delta_{aS} = \omega_{aS} - \omega_{aS}^0$ representing the detuning from the anti-Stokes frequency $\omega_{aS}^0$ defined by the carrier laser frequencies and $\Delta = \Omega - (\omega_1^0 - \omega_2^0)$ quantifying the ordinary Raman detuning from the vibrational frequency $\Omega$. It must be pointed out that Eq. (2) applies to an isolated line but the extension to multiple lines is simply realized by including the proportionality factor not appearing in Eq. (2) and by taking the summation of the corresponding line terms. Another general feature regards time aspects of the laser fields. As a matter of fact, Eq. (2) holds good for CARS polarization generated with laser pulses of whatever duration. However, similar to many experimental works and theoretical analysis, we take synchronous excitation of the Raman response (hence, $\tau_{12} = 0$ and $\tau_{23} = \tau$) and the time-delayed Yuratich equation for impulsive pump and Stokes pulses simplifies to

$$P_{CARS}^{(3)}(\omega_{aS},\tau) \propto \int_{-\infty}^{\infty} e^{-i\omega\tau} \frac{\mathcal{E}_3(\delta_{aS}-\omega)}{\Delta-\omega-i\Gamma/2} d\omega. \quad (3)$$

Understandably, in solving Eq (2) and especially Eq. (3), the frequency-domain envelope of the probe field assumes a key role and, for this reason, it is appropriate to specify the problem to some well-known probe shapes that, in some instances, are reckoned theoretically challenging within the time model of Eq. (1). In particular, in the following part of this Letter, we refer to four examples of Fourier-transform pairs for the probe field: Gaussian-Gaussian (GG) probe of frequency linewidth $\sigma$ [1, 13, 16, 18, 28], Exp-Lorentzian (EL) probe of frequency linewidth $\gamma$ [9, 12, 14, 18], Sinc-Square (SiSq) probe characterized by the spectral Rect function of width $W$ [2, 4, 6, 18] and its inverse Square-Sinc probe (SqSi) characterized by the spectral Sinc function associated with a time-domain square pulse of duration $t_{pr}$ [7]. These examples are summarized in Tab. 1 where closed-form solutions of Eq. (3) are shown and compared below to experimental cases taken from the literature.

Table 1. Probe fields and their corresponding third-order polarizations for CARS generation

| $\mathcal{E}_3(\omega)$ | $P_{CARS}^{(3)}(\omega_{aS},\tau)$ |
|---|---|
| $e^{-\omega^2/(2\sigma^2)}$ | $e^{-i\delta_{aS}\tau} e^{-(\sigma\tau)^2/2} w(\zeta)$ |
| $\dfrac{1}{(\gamma/2-i\omega)}$ | $\dfrac{\theta(\tau)e^{-(\Gamma/2+i\Delta)\tau}+\theta(-\tau)e^{(\gamma/2-i\delta_{aS})\tau}}{\Delta-\delta_{aS}-i(\Gamma+\gamma)/2}$ |
| $\text{Rect}_W(\omega)$ | $e^{-(\Gamma/2+i\Delta)\tau} \Lambda(\delta_{aS}-\Delta,\Gamma,\tau,W)$ |
| $\dfrac{t_{pr}}{2}\text{Sinc}(\omega t_{pr}/2)$ | $\dfrac{e^{-i\delta_{aS}\tau}}{z} S(\omega_{aS},\tau)$ |

Tab. 1 is complemented by the following definitions
$w(\zeta) = e^{-\zeta^2} Erfc(-i\zeta)$ (Faaddeva function),
$\zeta = \sqrt{2}\{(\delta_{aS}-\Delta)/\sigma + i[\Gamma/(2\sigma)-\sigma\tau]\}/2$,
$\Lambda(\delta_{aS}-\Delta,\Gamma,\tau,W) = Gamma[0,z_+,z_-] +$
$\qquad\qquad 2\pi i \theta(\tau)\text{Rect}_W(\delta_{aS}-\Delta)$,
$z_\pm = -[\Gamma/2-i(\delta_{aS}-\Delta\pm w/2)]\tau$,
$S(\omega_{aS},\tau) = Sign(1-2\tau/t_{pr}) + Sign(1+2\tau/t_{pr}) +$
$\qquad 2e^{2iz\tau/t_{pr}}[\theta(2\tau/t_{pr}-1)e^{-iz} - \theta(1+2\tau/t_{pr})e^{iz}]$
$z = t_{pr}(\delta_{aS}-\Delta+i\Gamma/2)/2$,

and the proof now continues by putting the results to the test. To begin, we focus our attention on the GG probe that was considered theoretically in the analysis of Compton et al. [28] and later followed by Stauffer et al. [18]. Different from the former authors (who provide an analytical result at zero delay only) and Stauffer et al. (whose approximated result neglects the contribution of the Faaddeva function) we underline the role of $w(\zeta)$ for time-domain measurements of the Raman linewidths [10, 13, 17, 29]. To this end, we need to draw attention to the following fundamental dependences of the CARS signal of an isolated line

$$S_{GG}(\omega_{aS},\tau) \propto e^{-(\delta_{aS}-\Delta)^2/\sigma^2} e^{-\Gamma\tau} | Erfc(-i\zeta) |^2$$

(4)

and it's crystal clear to see that the complementary error function *Erfc* introduces spectral and time corrections to both the Gaussian lineshape and the exponential decay. In particular, the on-resonance slope of $Log[S_{GG}(\omega_{aS},\tau)]$ is $-\Gamma$ for time delays that make $|Erfc(-i\zeta)|^2$ stationary, that is $\tau > [2\sqrt{2} + \Gamma/(2\sigma)]/\sigma$. The condition corresponds to delays greater than about 200 ps for a probe-pulse duration of 100 ps and such a threshold delay agrees well with the experimental observations [17, 29].

Extension to multiple lines is easy. An example is reported in Fig. 1 for time-resolved CARS microscopy of benzaldehyde whose signal is understood by numerically solving Eq. (1) in the work of Volkmer et al. [1].

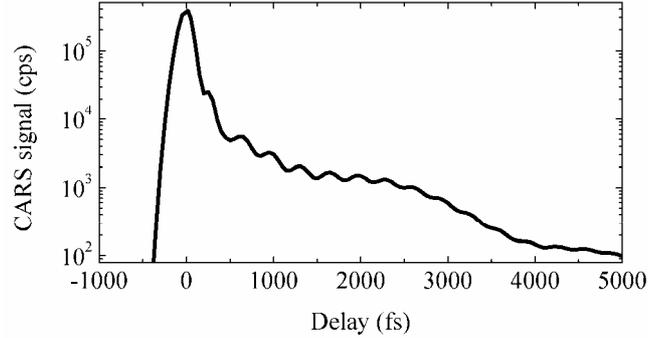

Fig. 1  Simulation of the time-domain CARS signal of benzaldehyde. The simulation includes the contribution of the non-resonant CARS component as well as the finite bandwidth of Gaussian pump and Stokes pulses. The result agrees with the measurements and the numerical solution of the time-dependent model reported by Volkmer et al. [1].

The simplest application of the time-delayed Yuratich equation is found for the EL probe. As shown in Tab. 1, the solution of Eq. (3) reproduces straightforwardly the Lorentzian spectral structure of the probe with time decay and rise controlled respectively by the Raman linewidth (for positive delays) and the laser linewidth (for negative delays). This case is very simple and has been analytically treated by others to a great extent [18].

On the other hand, the advantage of our approach over time-dependent calculations based on Eq. (1) [1-3, 6, 8, 11, 15, 18, 22] can be appreciated for the known case of a SiSq probe that is created by means of pulse shaping techniques [2, 4, 6, 18]. Latest research has made it clear that an analytical solution for the time-dependent CARS model can only be derived by expanding the time envelope of the probe field as a Taylor series about the probe delay $\tau$ [18]. By contrast, an analytical solution is actually achievable in terms of Gamma functions if the spectral approach is assumed (see Tab. 1). The formal demonstration is strengthened by the examples of Fig. 2 where CARS simulations for gas-phase $N_2$ and a solvated π-coniugated organic molecular species [*trans*-4-dimethylamino-4'-nitrostilbene (DANS) dissolved in acetonitrile (ACN)] are reported. The results are in excellent agreement with the experimental and numerical results of Staffer et al. [18].

To finalize our inspection of closed-form solutions of probe-shape effects in time-domain CARS, we examine the case of the Sq-Si probe [7]. This is analogous to the above-mentioned Si-Sq probe except for the exchange between the spectral and time variables in the Fourier pairs. The result of Tab. 1 for the Sinc spectral shape of $\varepsilon_3(\omega)$ can be further analyzed by discriminating CARS signals of different time delays (an example is given in Fig. 3 for $t_{pr} = 3$ ps). In brief, at negative delays $\tau < -t_{pr}/2$, the terms in the function $S(\omega_{aS},\tau)$ cancel out implying the impossibility of CARS generation as a consequence of time causality. At $|\tau| < t_{pr}/2$, the total CARS polarization is dominated by the non-resonant contribution (not shown in Tab. 1) that is constant in time but has a spectral shape dictated by the pump and Stokes fields (assumed for simplicity with Gaussian shapes). However, this featureless non-resonant polarization acts as a local oscillator that amplifies the resonant polarization of Tab. 1. The latter emerges without the non-resonant part at $\tau > t_{pr}/2$ and the signal is characterized by the usual exponential decay accompanied by a Sinc-type spectral depencence.

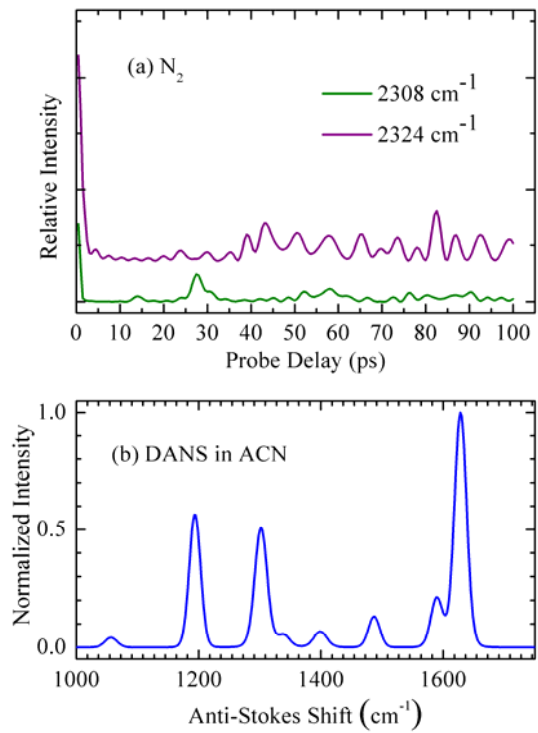

Fig. 2 (a) Time-domain N₂ signals for the $\text{Rect}_W(\omega)$ probe of Tab. 1 with $W$=12.2 cm⁻¹ and centered at 2308 and 2324 cm⁻¹. (b) DANS spectrum for the $\text{Rect}_W(\omega)$ probe of Tab. 1 with $W$=15.4 cm⁻¹.

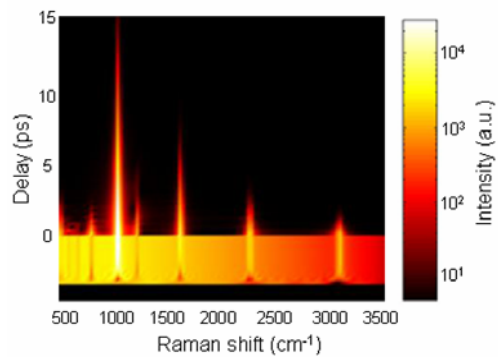

Fig. 3 Simulation of the time-domain spectrum of benzonitrile measured by Selm et al. [7].

In the end, given the foregoing reasoning and examples, we have proven that the time-delayed Yuratich equation offers a viable alternative to the more common time-dependent model of time-domain CARS [1-3, 6, 8, 11, 15, 18, 22]. One of the promising features of the suggested spectral approach consists in simple analytical solutions for probe shapes that are deemed impossible to be handled in a completely symbolic manner [18]. In this regard, other more complex laser shapes and pulse sequences are under study. In addition, practical examples are here made available in support of our argument. They provide convincing evidence that insights into time-domain CARS could be gained without the numerical effort that is often made in applications ranging from the more traditional gas-phase sensing down to the more recent biochemical microscopy.